\documentclass[10pt]{article}

\usepackage{epsfig}
\usepackage{wrapfig}
\usepackage{amsmath}
\usepackage{amssymb}
\usepackage{cite}

\begin{document}

\begin{figure*}[t]
\centering\epsfig{file=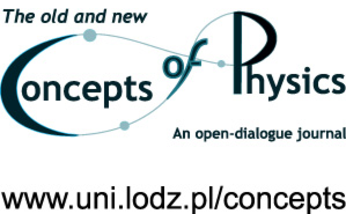,width=0.4\linewidth}
\end{figure*}


\begin{titlepage}

\title{\bf Time-to-Space Conversion in Neutrino Oscillations}

\author{\bf M.~I.~Shirokov$^*$ and V.~A.~Naumov$^{**}$ \\[1.5mm]
{\small Bogoliubov Laboratory of Theoretical Physics}  \\
{\small Joint Institute for Nuclear Research}          \\
{\small Joliot-Curie 6, RU-141980, Dubna, Russia}      \\[1mm]
{\small ${}^{ *}$e-mail: shirokov@theor.jinr.ru}       \\
{\small ${}^{**}$e-mail: vnaumov@theor.jinr.ru}
}

\date{\it\small (Received 25 July 2006; accepted 10 August 2006)}

\maketitle

\begin{abstract}

We study the neutrino oscillation problem in the framework of the wave packet formalism.
The neutrino state is described by a packet located initially in a region $S$ (source)
and detected in another region $D$ at a distance $R$ from $S$.
We examine how the oscillation probability as a function of variable $R$ can be derived
from he oscillation probability as a function of time $t$, the latter being found
by using the Schr\"odinger equation.
We justify the known prescription $t\to R/c$ without referring to a specific form of the
neutrino wave packet and only assuming the finiteness of its support.
The effect of the oscillation damping at large $R$ is revealed.
For an illustration, an explicit expression for the damping factor is obtained using
Gaussian packet.

\vspace*{5mm}


\end{abstract}

\thispagestyle{empty}

\end{titlepage}

\section{Introduction}
  \label{Introduction}

While in the minimal formulation of the standard electroweak model neutrinos
are massless,
there have been strong experimental evidences for nonzero neutrino masses and
mixing of neutrino flavors which fortify the idea of Pontecorvo~\cite{Pontecorvo}
that nonzero neutrino mass and lepton flavor violation would produce spontaneous
oscillation of neutrinos from one flavor to another in a manner similar to
the strangeness oscillations of neutral kaons.
The consistent quantum-mechanical treatment of neutrino oscillations is important
for interpretation of the current and future experiments.

A typical neutrino oscillation experiment can be described by the following scheme.
There is a neutrino source which produce neutrinos of a definite flavor
$\nu_\alpha$ ($\alpha=e,\mu,\tau$) in a finite region $S$.
The detector of neutrinos of the same or another flavor $\nu_\beta$ is located
in another finite region $D$ at a distance $R$ from $S$.
The regions $S$ and $D$ are supposed to be well separated from one another.
The probability to detect $\nu_\beta$ with $\beta\neq\alpha$ is in general nonzero.
The theoretical task is to predict the $R$ dependence of this probability.
For simplicity, only the neutrino oscillations in vacuum are considered here.

There are many approaches to this problem. The simplest is to find first
the temporal evolution of the neutrino state and then to convert it into
the spatial evolution. Usually it is assumed that the neutrino state at $t=0$
has a definite momentum that is described by a plane wave. Then, by applying
the Schr\"odinger equation, one finds the probability $W(t)$ to detect the same
neutrino flavor at $t>0$ (survival probability) or another flavor (transition
probability) at $t>0$ (Sect.~\ref{Time}).
In order to convert the time dependence of $W$ into the $R$ dependence,
one apply the \emph{classical propagation prescription} (in the terminology of
Ref.~\cite{Beuthe:01}), according to which $t$ is simply replaced by $R/c$
(where $c=1$ is the velocity of light), 
with $R$ being displacement of the classical ultrarelativistic particle at time $t$.
The deficiencies of this prescription are well known
(see, e.g., Ref.~\cite[Ch.~2.3.3]{Rich:93}).
In particular, the formula $R=vt$ is only valid for a classical particle
          trajectory and is inconsistent with quantum postulates.
          
It is generally accepted that one must describe the neutrino state by a wave
packet rather than by a plane wave. The relevant considerations are called
in Ref.~\cite{Beuthe:01} \emph{intermediate wave packet models}. Many aspects of
of these models were discussed in
Refs.~\cite{Kayser:81,Giunti:91,Giunti:92,Giunti:97,Nauenberg:98,Lipkin:99,%
            Takeuchi:99,Takeuchi:00,Giunti:03a,Giunti:03b,DeLeo:04,%
            Bernardini:04,Bernardini:05,Asahara:05}.
The present paper belongs to this class.
Some peculiarities of our approach are presented in the concluding section
(Sect.~\ref{Conclusion}).
One of our aims (in a sense, pedagogical) is to propose an example 
of how one can obtain the time-to-space conversion within the framework of the quantum
postulates.

In other formalisms, like the \emph{external wave packet models} and
\emph{quantum field theoretical} or \emph{$S$ matrix approach}
(see Ref.~\cite{Beuthe:01} and references therein), the wave packet description
is used for the particles which either create neutrino (e.g., a pion,
when neutrinos are produced through the $\pi_{\mu2}$ decay) or detect
it (e.g., an oxygen nucleus in a water-Cherenkov detector). Being more advanced,
these approaches are more cumbersome and usually are aimed at experiments of a
specific type.
The intermediate wave packet models are simpler and, in a sense, more general,
being applicable for generic neutrino source and detector.
What is more, they are sufficient for the purpose of the present study: to show how the
flavor oscillations in time leads to these in space (Sect.~\ref{Space}).

Our main results are summarized in Sect.~\ref{Conclusion}.

\section{Neutrino oscillations in time}
  \label{Time}

In order to define what ``neutrino oscillations in time'' means and to
prepare equations necessary for the next Section, let us discuss the
following problem (c.f., for example, Ref.~\cite{Rich:93}).
One free electron neutrino state with definite momentum $\mathbf{k}$
(\emph{weak eigenstate}) is prepared at the moment $t=0$.
Let $\Psi(0)\equiv|e,\mathbf{k}\rangle$ denotes this initial state
(as usual, we omit spin indices). The evolution of the state in time is
governed by a time-independent and nondiagonal in the flavor basis Hamiltonian $\hat{H}$,
the \emph{propagation Hamiltonian}, according to Ref.~\cite[Ch.~2]{Beuthe:01}.
Then $\Psi(t)=\exp\left(-i\hat{H}t\right)\Psi(0)$. According to the
Pontecorvo's supposition~\cite{Pontecorvo}, $\hat{H}$ has eigenvectors
$|j,\mathbf{k}\rangle$ ($j=1,2,\ldots$) with definite value of the total
momentum $\mathbf{k}$ (\emph{mass eigenstates}). The corresponding
eigenvalues are equal to $m_j$ at $\mathbf{k}=0$ and to
$E_j=\sqrt{|\mathbf{k}|^2+m_j^2}$ at $\mathbf{k}\ne0$.
The weak eigenstate $\Psi(0)$ may be expanded in terms of the mass eigenstates
(below, we assume $j=1,2$ for simplicity):
\begin{equation}\label{PontecorvoRule}
|e,\mathbf{k}\rangle=\cos\theta|1,\mathbf{k}\rangle
                    +\sin\theta|2,\mathbf{k}\rangle.
\end{equation}
Then
\begin{equation}\label{PontecorvoRule2}
\Psi(t)=|e,\mathbf{k};t\rangle
       =e^{-iE_1t}\cos\theta|1,\mathbf{k}\rangle
       +e^{-iE_2t}\sin\theta|2,\mathbf{k}\rangle.
\end{equation}
This equation provides all information on the neutrino state at time $t$
(expectation values of observables, probabilities).
In particular, the probability of finding the system in the initial state
(survival probability) is given by
\begin{equation}\label{SurvivalProbability}
W_{\mathbf{k}}(t)= \left|\langle e,\mathbf{k}|e,\mathbf{k};t\rangle\right|^2
=1-\frac{1}{2}\sin^22\theta\left[1-\cos(E_2-E_1)t\right].
\end{equation}
For ultrarelativistic neutrinos ($|\mathbf{k}|^2 \gg m_{1,2}^2$),
Eq.~\eqref{SurvivalProbability} is simplified to
\begin{equation}\label{SurvivalProbability_UR}
W_{\mathbf{k}}(t)\approx1-\frac{1}{2}\sin^22\theta\left[1-
\cos\left(\frac{2\pi t}{T}\right)\right]
\end{equation}
with
\begin{equation*}\label{T_osc}
T=\frac{4\pi|\mathbf{k}|}{\left|m_2^2-m_1^2\right|}.
\end{equation*}
This probability is a periodic function of time (with the period
$T$) and provides the simplest example of \emph{neutrino oscillations}
in time. Replacing $t$ by $R/c$ in Eq.~\eqref{SurvivalProbability_UR} one obtains the
``standard'' formula for the ultrarelativistic neutrino oscillations in space:
\begin{equation}\label{SurvivalProbability_UR_Space}
W_{\mathbf{k}}(R/c)\approx1-\frac{1}{2}\sin^22\theta\left[1-
\cos\left(\frac{2\pi R}{\Lambda}\right)\right]
\end{equation}
with
\begin{equation*}\label{Lambda_osc}
\Lambda=Tc.
\end{equation*}

\paragraph{Note~1.}

For deriving Eq.~\eqref{SurvivalProbability} we imply the unit normalization
of the mass eigenstates: 
\begin{equation*}
\langle i,\mathbf{k}|j,\mathbf{k}\rangle=\delta_{ij}.
\end{equation*}
Since these states are the eigenvectors of the momentum operator $\hat{P}$, this assumption
means that the operator $\hat{P}$ has a discrete spectrum. This assumes that the system under
consideration is in a large spatial volume and the usual periodicity conditions are imposed
(or the opposite boundaries of the volume are identified). However, in what follows we will
approximate the summations over $\mathbf{k}$ by integrations over the 3-momentum.
 
\paragraph{Note~2.}

Let us consider the muon neutrino created in the decay $\pi^+\to\mu^++\nu_\mu$,
$\pi^+$ being at rest. From the energy-momentum conservation it follows that
the neutrino state is a superposition of the plane waves $|\mathbf{k}_j\rangle$,
$j=1,2,\ldots,$ with the $|\mathbf{k}_j|$ unambiguously defined by the masses $m_\pi$,
$m_\mu$ and $m_j$ (see, e.g., Ref.~\cite[Sect.~3]{Giunti:00}).
In contrast, the ``Pontecorvo's prescription''
\eqref{PontecorvoRule} assumes that the neutrino initial state is a single
plane wave with an arbitrary momentum.
Here we accept for simplicity just this prescription.

\section{Neutrino oscillations in space}
  \label{Space}

In any neutrino oscillation experiment the source of neutrinos with definite
flavors is localized in a finite volume $S$ and the detector of neutrinos
(of the same or different flavor) is localized in another finite region $D$
placed at a distance $R$ from $S$. The source cannot prepare a neutrino with
definite momentum $\mathbf{k}$ (plane wave). The emitted neutrino
(e.g., electron neutrino) must be described by a packet, i.e.\ by a superposition of
the states $|e,\mathbf{k}\rangle$, $\forall\mathbf{k}$ which form a complete
set of states:
\begin{equation}\label{Packet}
|e,g\rangle=\int d^3k\tilde{g}(\mathbf{k})|e,\mathbf{k}\rangle,
\quad
\int d^3k|\tilde{g}(\mathbf{k})|^2=1.
\end{equation}
In the coordinate representation, the state $|e,g\rangle$ is described by the function
\begin{equation}\label{Packet_x}
g(\mathbf{x})=\langle e,\mathbf{x}|e,g\rangle
=(2\pi)^{-3/2}\int d^3k\tilde{g}(\mathbf{k})\exp\left(i\mathbf{kx}\right),
\end{equation}
which is the amplitude of the density probability  that electron neutrino in the state 
$|e,g\rangle$ has the position $\mathbf{x}$.
It is assumed that $g(\mathbf{x})$ is localized in $S$
in the sense that $g(\mathbf{x})$ is negligibly
small (physically indistinguishable from zero) outside $S$.

After multiplying both sides of Eq.~\eqref{PontecorvoRule} by $\tilde{g}(\mathbf{k})$ and
integrating over $\mathbf{k}$ we obtain the expansion of $|e,g\rangle$ over the mass
eigenstates:
\begin{equation*}\label{eg_expansion_1}
|e,g\rangle=\cos\theta\int d^3k\tilde{g}(\mathbf{k})|1,\mathbf{k}\rangle
           +\sin\theta\int d^3k\tilde{g}(\mathbf{k})|2,\mathbf{k}\rangle.
\end{equation*}
Analogously, from Eq.~\eqref{PontecorvoRule2} one gets
\begin{align*}\label{eg_expansion_2}
|e,g;t\rangle
& = \exp\left(-i\hat{H}t\right)|e,g\rangle \nonumber \\
& = \cos\theta\int d^3k\tilde{g}(\mathbf{k})e^{-iE_1t}|1,\mathbf{k}\rangle
    +\sin\theta\int d^3k\tilde{g}(\mathbf{k})e^{-iE_2t}|2,\mathbf{k}\rangle,
\end{align*}
where
\begin{equation*}\label{Energies}
E_j=\sqrt{|\mathbf{k}|^2+m_j^2}.
\end{equation*}
For the amplitude to find neutrino $\nu_e$ in the point $\mathbf{x}$ we obtain
\begin{equation*}\label{Amplitude}
\langle e,\mathbf{x}|e,g;t\rangle=\cos^2\theta g_1(\mathbf{x},t)
                                 +\sin^2\theta g_2(\mathbf{x},t),
\end{equation*}
\begin{equation}\label{Amplitude_t}
g_j(\mathbf{x},t)=(2\pi)^{-3/2}\int d^3k\tilde{g}(\mathbf{k})
\exp\left[-i\left(E_jt-\mathbf{kx}\right)\right].
\end{equation}
Here we have used the equalities
\[
\langle e,\mathbf{x}|1,\mathbf{k}\rangle=\cos\theta e^{i\mathbf{kx}},
\quad
\langle e,\mathbf{x}|2,\mathbf{k}\rangle=\sin\theta e^{i\mathbf{kx}}.
\]
The functions $g_1(\mathbf{x},t)$ and $g_2(\mathbf{x},t)$ describe the evolving wave
packets of particles with masses $m_1$ and $m_2$, respectively. According to
Eqs.~\eqref{Packet} and \eqref{Amplitude_t}, they coincide with the packet $g(\mathbf{x})$,
Eq.~\eqref{Packet_x}, at the initial moment $t=0$.
It is shown in Appendix~A that, neglecting the spread of the packet, the evolution
of $g_j(\mathbf{x},t)$ reduces, up to a phase factor (close to 1 in the ultrarelativistic case),
to the ``drift'' of the initial packet $g(\mathbf{x})$ with the velocity
$\mathbf{v}_j=\mathbf{k}/E_j$:
\[
\left|g_j(\mathbf{x},t)\right|=\left|g(\mathbf{x}-\mathbf{v}_jt)\right|.
\]
In other words, the wave packets $g_j(\mathbf{x},t)$ approximately follow the
trajectory of a classical  particle.

Let us now assume that the detector $D$ registers the electron neutrinos. The probability
to find in the state $|e,g;t\rangle$ one electron neutrino localized within the region $D$
is equal to
\begin{align}\label{W_g1}
W_g(t)& = \int_Dd^3x \left|\langle e,\mathbf{x}|e,g;t\rangle\right|^2 \nonumber \\
      & = \int_Dd^3x\left\{
           \cos^4\theta\left|g_1(\mathbf{x},t)\right|^2
          +\sin^4\theta\left|g_2(\mathbf{x},t)\right|^2\right.        \nonumber \\
      &   \left.+2\cos^2\theta\sin^2\theta\,
           \text{Re}\left[g_1^*(\mathbf{x},t)g_2(\mathbf{x},t)\right]
\right\}.
\end{align}
This probability may also be represented through the average value of the projection operator
\[
\hat{T}_D=\int_Dd^3x|e,\mathbf{x}\rangle\langle e,\mathbf{x}|,
\]
namely
\[
W_g(t)=\left|\langle e,g;t|\hat{T}_D|e,g;t\rangle\right|^2.
\]

\begin{figure}[ht]
\epsfig{file=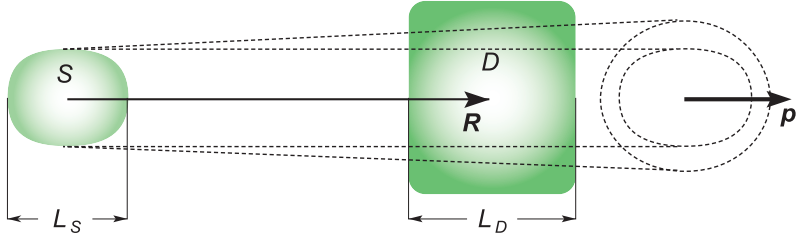,width=\linewidth}
\caption{The mutual disposition of the neutrino source $S$ and neutrino
         detector $D$. The dimensions of the source and detector and the distance
         $R$ between $S$ and $D$ are implied to satisfy the inequalities
         $L_S\ll L_D \ll R$. The space between dashed lines is covered by the supports
         of moving wave packet as it shifts and (slowly) spreads. The vector $\mathbf{p}$
         is the average momentum of the packet.}
\label{Fig:SD}
\end{figure}

We consider now the behavior of $W_g$ as a function of $t$. Let $\mathbf{R}$
be the vector connecting the center of $S$ with the center of $D$ ($R=|\mathbf{R}|$
being the distance between $S$ and $D$), see Fig.~\ref{Fig:SD}.
Suppose that the module $p$ of the average momentum of the packet is much greater than
$m_1$ and $m_2$ (the neutrino are ultrarelativistic) and therefore the group velocities
$v_1$ and $v_2$ are both approximately equal to the velocity of light.

\begin{wrapfigure}{r}{0.54\linewidth}
\epsfig{file=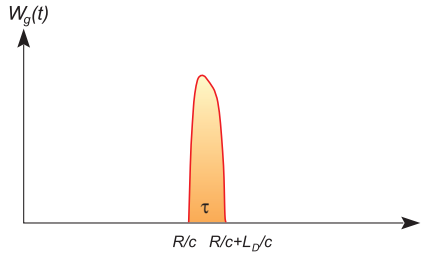,width=\linewidth}
\caption{The probability of electron neutrino localization in the
         detector region as a function of time.}
\label{Fig:Pulse}
\end{wrapfigure}

So, the packets $g_1(\mathbf{x},t)$ and
$g_2(\mathbf{x},t)$ enter the region $D$ (and leave it) almost simultaneously.
Their nonoverlapping will be discussed later on.
Till the moment $t=R/c$, the moving supports of the packets $g_1$ and $g_2$ do not cover the
region $D$ ($\text{supp}(g_j)\cap D =\emptyset$) and therefore $W_g(t)=0$ if $t<R/c$.
After the moment $(R+L_D)/c$ (where $L_D$ is the dimension of the region $D$ in the direction
of the vector $\mathbf{R}$) the packets get out of $D$ and we have $W_g(t)=0$ again
(see Fig.~\ref{Fig:Pulse}).
Let us assume further that 
\begin{equation}\label{Conditions}
L_S\ll L_D\ll R, 
\end{equation}
where $L_S$ is the dimension of the region $S$ (which may be considered as coinciding
with the dimension of the initial packet) along the vector $\mathbf{R}$
(see Fig.~\ref{Fig:SD}).
Then we can neglect all marginal effects, e.g.\ the gradual growing of $W_g(t)$
as neutrino packets are entering the detector.
So we obtain that $W_g(t)$ is nonzero
only within the time interval $\tau=(R/c,R/c+L_D/c)$. Inside this interval, the probability
$W_g(t)$ is approximately equal to $W_g(R/c)$, $L_D$ being much smaller than $R$.
Note that if the packet spreading is taken into account, one cannot consider
arbitrary large values of $R$ since the support of the spread packet must not exceed $D$.

It is sometimes convenient to deal with the number of particles rather than the probabilities.
Consider an ensemble of $N_0$ neutrinos in the initial state $|e,g\rangle$.
Then the total number $N$ of neutrinos $\nu_e$ falling into the detector at all times from
$t=0$ to $\infty$ is proportional to $N_0W_g(R/c)$.
Being independent of time, this number depends upon $R$. 
Since the neutrinos have a nonzero average momentum $\mathbf{p}$ then, during the
time $\tau$ there is a nonzero current of neutrinos in $D$.
For example, we can consider a detector which uses the Pontecorvo's reaction
\begin{equation}\label{PontecorvoReaction}
\nu_e+\text{Cl}\to\text{Ar}+e^-
\end{equation}
for catching the electron neutrinos (see, e.g., Ref.~\cite[Ch.~2.3.3]{Rich:87}).
The number of the resulting Ar nuclei depends on the number $N$ of electron neutrinos
falling into the detector, density of the chlorine nuclei in $D$, total cross section
of the reaction \eqref{PontecorvoReaction}, etc.
However, all these factors but $N$ do not depend upon $R$.
We may assume that all detectors placed at different distances $R$ from $S$ are identical.
It is pertinent to remark that, assuming the inequalities \eqref{Conditions}, we imply that
each neutrino detected in the chlorine detector had been actually emitted by a small piece
of the Sun (not by the Sun as a whole). The total number of neutrinos detected in the
chlorine experiment must be confronted to the probability $W_g(R/c)$, Eq.~\eqref{W_g1},
summed over all these small pieces which make up the Sun.

Let us remind that the chlorine experiment does not register the time when Ar nuclei appear
while measures the number of the argon nuclei accumulated in the detector during a long
period.
Thus, instead of the dependence on time there appears the dependence on $R$ (through $N$).
If $R$ is fixed then the actual time of appearing of Ar nuclei in $D$ is also fixed: $t=R/c$.
According to Beuthe~\cite[Ch.~2.3.1]{Beuthe:01}, this is called \emph{conversion of the
evolution in time into the evolution in space}. This conversion has been obtained above
by using the wave packets but their specific form was not employed (only finiteness of their
supports was used).

Our next task is to calculate numerically the probability $W_g$, Eq.~\eqref{W_g1} in
order to ascertain if one can obtain from  $W_g$ the standard formula
\eqref{SurvivalProbability_UR_Space} for $W_{\mathbf{k}}(t)$.
This calculation reduces to the calculation of the integrals
\begin{equation}\label{I_j}
\int_Dd^3x \left|g_j(\mathbf{x},t)\right|^2, \quad j=1,2
\end{equation}
and
\begin{equation}\label{I_12}
\int_Dd^3x g_1^*(\mathbf{x},t)g_2(\mathbf{x},t) \equiv I_{12}(t).
\end{equation}
Let us assume that the packets $g_1(\mathbf{x},t)$ and
$g_2(\mathbf{x},t)$ are entirely inside $D$ at $t\in\tau$, $\tau\equiv(R/c,R/c+L_D/c)$
(see note below). Then the integrals \eqref{I_j} can be replaced by the integrals over
the whole three-dimensional space. The latter do not depend upon $t$, both being equal
to the normalization integral 
\begin{equation}\label{Normalization}
\int d^3x \left|g(\mathbf{x})\right|^2=1.
\end{equation}

\paragraph*{Note.}

Besides shifting in space, the packets $g_i(\mathbf{x},t)$ also spread
(see Appendix~A). There is a boundary value $R_{\text{spr}}$ of the distance $R$ such
that the spread packets completely cover the region $D$ when $R>R_{\text{spr}}$.
Then the integrals \eqref{I_j} are not equal to 1. If $R<R_{\text{spr}}$ then the
packet's supports are in $D$. One may define $R_{\text{spr}}$ as follows.
Let $g_i(\mathbf{x},t)$ are Gaussian, see Eq.~\eqref{A4} or 
Eq.~\eqref{GaussianPacket}. Then its transversal spatial dispersions equal to
$\sigma_{1,2}^2(t)$, see
Eq.~\eqref{A8}, its spacial dimension being $\sigma_{1,2}(t)$. The dimension is equal to
$L_D$ (the region $D$ dimension) when
$t \sim t_{\text{spr}}=R_{\text{spr}}/c=2L_D\sigma p$
(Eq.~\eqref{A8} is used and inequality $L_D \gg L_S \sim \sigma$ is presumed).
Here $\sigma$ is the spatial dimension of the initial packet $g(\mathbf{x},0)$ and
$p$ is the magnitude of the average momentum of the packet.
So integrals \eqref{I_j} equal to 1 if $R<R_{\text{spr}}=2 L_D\sigma p c$.
If, for example, $g$ is a macroscopical packet (say $\sigma \sim1$~cm) with the
average energy $E \simeq p$ of the order of several MeV then
$R_{\text{spr}}\sim10^{12}L_D$.

The integral $I_{12}$ defined by Eq.~\eqref{I_12} is more complicated.
It approximately equals to 1 when $t$ is small (along with $R$;
remind that $t\in\tau$ with $\tau\equiv(R/c,R/c+L_D/c)$)
because $g_1(\mathbf{x},t) \simeq g_2(\mathbf{x},t)$ at small $t$.
The integral $I_{12}$ vanishes
when $t\sim R/c$ is sufficiently large because $g_1$ and $g_2$ cease to overlap
(being both in $D$ if $R<R_{\text{spr}}$) due to different group velocities,
$v_1 \neq v_2$. For such $R$ the probability $W_g(R/c)$ does not longer depend on $R$
and does not oscillate.
This phenomenon does not exist in the standard formula \eqref{SurvivalProbability_UR_Space}.

For a numerical illustration we calculated in Appendix~B the integral $I_{12}$ by using
the Gaussian packet
\begin{equation}\label{GaussianPacket}
\tilde{g}(\mathbf{k}) \propto
\exp\left[-\left(\mathbf{p}-\mathbf{k}\right)^2\sigma^2\right].
\end{equation}
Here $\mathbf{p}$ is the average momentum and $\sigma^2$ is the spatial dispersion of the
packet. It is assumed that $p\equiv|\mathbf{p}|\gg m_i$ and that
$\sigma$ is macroscopically large, so that we have
\begin{equation}\label{MacroConditions}
p\sigma\gg1.
\end{equation}
We suppose that $R\ll R_{\text{spr}}\sim L_D \sigma pc$ so that the integrals 
\eqref{I_j} are equal to 1.
The resulting value of $W_g(t)$ at the moment $t=R/c$ can be represented as
\begin{equation}\label{W_g2}
W_g(R/c)=1-\frac{1}{2}\sin^22\theta\left[1-\mathcal{D}
\cos\left(\frac{2\pi R}{\Lambda}\right)\right],
\end{equation}
where
\begin{equation}\label{D}
\mathcal{D}=\exp\left[-\left(\frac{R}{\Lambda}\right)^2
\left(\frac{\pi}{|\mathbf{p}|\sigma}\right)^2\right]
\end{equation}
and
\begin{equation*}\label{Lambda}
\Lambda=Tc=\frac{4\pi|\mathbf{p}|}{\Delta m^2},
\quad
\Delta m^2=|m_1^2-m_2^2|.
\end{equation*}
So, the spatial oscillations of $W_g$ are damped by the factor $\mathcal{D}$.
The damping can be neglected when $R$ is such that $\mathcal{D}\simeq1$ and it is
significant when $R$ is such that $\mathcal{D}<1$.
Let us designate the corresponding boundary value of $R$ as $R_{\text{damp}}$ and
define it as that value of $R$ at which $\mathcal{D}\sim1/e$.
Then, as it is seen from Eq.~\eqref{D}, $R_{\text{damp}}\sim\Lambda\sigma pc$ and,
according to the inequality \eqref{MacroConditions}, $R_{\text{damp}}\gg\Lambda$.
So the damping can be neglected, e.g., for $R$ of the order of one or several
oscillation length $\Lambda$.
In this case, $W_g(R/c)$ turns into the r.h.s. of Eq.~\eqref{SurvivalProbability_UR_Space}
for $W_{\mathbf{k}}(R/c)$, i.e., into the standard neutrino oscillation formula.

Now, let us compare $R_{\text{damp}}$ with $R_{\text{spr}}$. We have
\begin{equation}\label{dump-spr}
\frac{R_{\text{damp}}}{R_{\text{spr}}} \sim
\frac{\Lambda}{L_D} \sim
\frac{p}{\Delta m^2}\cdot\frac{1}{L_D} =
\frac{p}{\sqrt{\Delta m^2}}\cdot\frac{1}{L_D\sqrt{\Delta m^2}}.
\end{equation}
Due to the inequality $p\gg m_i$ the first factor in r.h.s. of Eq.~\eqref{dump-spr}
is large, but $L_D\sqrt{\Delta m^2}$ can be estimated as being much larger and
therefore $R_{\text{damp}}<R_{\text{spr}}$.

Thus we have obtained the following. When $R\ll R_{\text{damp}}$ the standard
formula is correct. As $R$ grows and becomes $\gtrsim R_{\text{damp}}$, the damping turns
out to be essential and the standard formula fails, while our result, given by
Eq.~\eqref{W_g1} still holds.
For very large $R>R_{\text{spr}}$, our formula becomes also incorrect due to the packet
spreading.
We conclude therefore that there exists an interval of $R$ values for which our formula
\eqref{W_g1} (containing the damping) is valid, while the standard formula is incorrect.

\section{Conclusion}
  \label{Conclusion}

The above consideration of neutrino oscillations is of pedagogical interest since
it does not use the reasons inconsistent with quantum postulates.
In particular, we have presented a purely quantum-mechanical justification of the
``classical propagation prescription'' $t \longmapsto R/c$. Usually, this prescription
is grounded on the classical reasons inconsistent with quantum mechanics.

Our justification is based on the observation that fixing the detector at a distance $R$
from the neutrino source $S$ automatically fixes the neutrino sojourn time in the detector
volume.
The standard result given by Eq.~\eqref{SurvivalProbability_UR_Space} for the spatial
neutrino oscillations follows from Eq.~\eqref{W_g1} when the spreading of the neutrino
packet is neglected. 
It is shown that the spreading cannot be neglected if $R$ is too large.
In this case the dimensions of the spread neutrino packets become larger than
the detector dimension $L_D$: the packets cover the detector.
An explicit estimation $R>R_{\text{spr}}$ is obtained by using a specific (Gaussian)
form of the packet (the meaning of $R_{\text{spr}}$ is explained in the Note after
Eq.~\eqref{Normalization}, Sect.~\ref{Space}).

We point out another qualitative reason for the breakdown of the standard formula.
Consider, for example, the case when dimensions of the packets $g_1(\mathbf{x},t)$ and
$g_2(\mathbf{x},t)$ of neutrinos with masses $m_1$ and $m_2$ are both less then $L_D$
(both being in $D$) but their supports do not overlap due to different group velocities
$v_1 \neq v_2$ (see text before Eq.~\eqref{GaussianPacket}). Then the oscillations cease.
The explicit example with the Gaussian packets shows that in the case of incomplete
overlapping there arises a damping of oscillations which is significant for large $R$.
($R \sim R_{\text{damp}}\ll R_{\text{spr}}$; for $R_{\text{damp}}$ see the text before
Eq.~\eqref{dump-spr}).
The resulting formula for the spatial oscillations reduces to the standard one for
$R \ll R_{\text{damp}}$ when the damping can be neglected.

\clearpage

\section*{Appendix~A: Packet evolution}
\label{A}

Let us estimate the integral \eqref{Amplitude_t}. Let the function 
$\tilde{g}(\mathbf{k})$ in Eq.~\eqref{Amplitude_t} be negligibly small
outside the sphere $|\mathbf{k}-\mathbf{p}|^2\le l_p^2$ with the center $\mathbf{p}$
and radius $l_p$ in the momentum space. We may choose $l_p$ to be several times larger
than the dispersion of the momentum distribution $\tilde{g}(\mathbf{k})$.
Assume that $\tilde{g}(\mathbf{k})=\tilde{S}(\mathbf{p}-\mathbf{k})$,
where $\tilde{S}(\mathbf{k})$ is a spherically symmetric function which is small
when $|\mathbf{k}|^2 > l_p^2$. The average momentum
$\int d^3k\tilde{g}(\mathbf{k})\mathbf{k}$ is equal to $\mathbf{p}$.

Having in mind (ultra)relativistic neutrinos one can safely assume that $p\gg m_i$.
Nevertheless, for generality, we leave $p$ arbitrary.
The integral \eqref{Amplitude_t} will be estimated for the case
$l_p\ll E=\sqrt{\mathbf{p}^2+m^2}$. The corresponding spatial dimension $l_x$ of the
initial packet $g(\mathbf{x},0)$ satisfies the inequality $l_x\gg1/E$.
In particular, $l_x$ may be macroscopically large; e.g., for neutrino energies
of the order of several MeV it can be as large as 1~cm.

Substituting $\tilde{g}(\mathbf{k})=\tilde{S}(\mathbf{p}-\mathbf{k})$ into
Eq~.\eqref{Amplitude_t} and changing the variable $\mathbf{k}\to\mathbf{p}-\mathbf{k}$
we arrive at the following result (index $j$ is omitted for simplicity)
\begin{align}\label{A1}
g(\mathbf{x},t)
& = (2\pi)^{-3/2}\int d^3k\exp(i\mathbf{kx})\tilde{S}(\mathbf{p}-\mathbf{k})
    \exp\left(-it\sqrt{|\mathbf{k}|^2+m^2}\right)                          \nonumber \\
& = \frac{\exp(i\mathbf{px})}{(2\pi)^{3/2}}\int d^3k\tilde{S}(\mathbf{k})
    \exp\left[-i\left(\mathbf{kx}+t\sqrt{|\mathbf{p}-\mathbf{k}|^2+m^2}\right)\right].
\end{align}
Since $\tilde{S}(\mathbf{k})$ does not vanish only when $|\mathbf{k}|\lesssim l_p$,
we may expand $\sqrt{|\mathbf{p}-\mathbf{k}|^2+m^2}$ in powers of $\mathbf{k}/E$:
\begin{equation}\label{A2}
\sqrt{|\mathbf{p}-\mathbf{k}|^2+m^2}
=E\left[1-\frac{\mathbf{pk}}{E^2}
+\frac{|\mathbf{k}|^2-\left(\mathbf{vk}\right)^2}{2E^2}+\ldots\right],
\quad \mathbf{v}=\frac{\mathbf{p}}{E}.
\end{equation}
Retaining only the two leading terms in the square bracket of Eq.~\eqref{A2} we obtain
\begin{align}\label{A3}
g(\mathbf{x},t)
&\simeq(2\pi)^{-3/2}\exp\left[i(\mathbf{px}-Et)\right]
                    \int d^3k\tilde{S}(\mathbf{k})
 \exp\left[-i\mathbf{k}\left(\mathbf{x}-t\mathbf{v}\right)\right] \nonumber\\
& = e^{-itm^2/E} g(\mathbf{x}-\mathbf{v}t).
\end{align}
Here $g(\mathbf{x})=g(\mathbf{x},0)$ is the initial packet.
Eq.~\eqref{A3} means that the packet evolution in this approximation
reduces to the shift of the initial packet; the phase factor
$e^{-itm^2/E}$ is inessential.

The additional spreading of the packet originates from other terms in the
square bracket of Eq.~\eqref{A2}. To give an idea of their effect on the packet
evolution we shall take into account the third term in the bracket assuming that
the momentum $\mathbf{p}$ is directed along $\mathbf{R}$ that is $\mathbf{p}=(0,0,vE)$.
Moreover, let us consider the particular case when the distribution $\tilde{S}(\mathbf{k})$
is Gaussian:
\begin{equation}\label{A4}
\tilde{S}(\mathbf{k})
=A\exp\left(-|\mathbf{k}|^2\sigma^2\right)
=A\exp\left[-\sigma^2(k_1^2+k_2^2+k_3^2)\right].
\end{equation}
Here $A$ is a constant and $\sigma^2$ is the spatial dispersion of the packet.
Then $g(\mathbf{x},t)$ reduces to the product of three integrals:
\begin{equation*}\label{A5}
g(\mathbf{x},t)=A\exp\left[i\left(\mathbf{px}-Et\right)\right]I_1I_2I_3.
\end{equation*}
Here
\begin{equation*}\label{AI_j}
I_j =\int_{-\infty}^{+\infty}d\kappa e^{-a_j\kappa^2-i\kappa\left(x_j-v_jt\right)};
\end{equation*}
\begin{equation*}\label{Av_j}
v_1=v_2=0,
\quad
v_3=v;
\end{equation*}
\begin{equation*}\label{a_j}
a_1=a_2=\sigma^2+\frac{it}{2E},
\quad
a_3=\sigma^2+\frac{it(1-v^2)}{2E}.
\end{equation*}
Note that $a_j$ are complex numbers with positive real parts.
According to Ref.~\cite[Ch.~2.5.36.1]{Prudnikov:81} the integrals $I_j$ are%
\begin{equation*}\label{A6}
I_j=2\int_0^\infty d\kappa e^{-a_j\kappa^2}\cos\kappa\left(x_j-v_jt\right)=
\sqrt{\frac{\pi}{a_j}}\exp\left[-\frac{\left(x_j-v_jt\right)^2}{4a_j}\right].
\end{equation*}
Hence we obtain
\begin{equation*}\label{A7}
|g(\mathbf{x},t)|^2=|AI_1I_2I_3|^2
=\text{Const}\cdot
\exp\left[-\sum_j\frac{(x_j-v_jt)^2}{2\sigma_j^2(t)}\right],
\end{equation*}
where
\begin{equation}\label{A8}
\sigma_1^2(t)=\sigma_2^2(t)=\sigma^2+\frac{t^2}{4\sigma^2E^2},
\quad
\sigma_3^2(t)=\sigma^2+\frac{t^2(1-v^2)}{4\sigma^2E^2},
\end{equation}
and the constant can be determined from the initial condition
and the normalization condition \eqref{Normalization} for the function $g(\mathbf{x})$.

The packet spatial dispersions along the axes $x_1$, $x_2$ and $x_3$ are all equal to
$\sigma^2$ at the initial moment, $t=0$, and increase with increasing time according
to Eq.~\eqref{A8}. In the ultrarelativistic case ($1-v^2\ll1$) and at large enough
$t$, $\sigma_{1,2}^2(t)$ may be much larger than $\sigma_3^2(t)$.
Taking into account that, according to our assumption, $\sigma$ is macroscopically
large ($\sigma\gg 1/p$), one can also conclude that increase in the packet
dimension in all directions is much less than the spatial shift $vt$ of the packet
as a whole.
Some numerical examples when the packet spreading can be neglected are presented in
Ref.~\cite[Ch.~3.1]{Goldberger:67}.
The physical conditions at which the spreading can be neglected in our task are examined
in  Sect.~\ref{Space}. 


\section*{Appendix~B: Evaluation of the integral $I_{12}$}
\label{B}

Let us evaluate the integral \eqref{I_12} defined in Sect.~\ref{Space}, for a practically
important particular case. Assume $t\in\tau$ and $R<R_{\text{spr}}$ (with $R_{\text{spr}}$
defined in Sect.~\ref{Space}) so that the supports of both $g_1(\mathbf{x},t)$ and
$g_2(\mathbf{x},t)$  are in $D$.
Hence the integration over the region $D$ in \eqref{I_12} can be extended to the whole
space. Now let us use the same assumption about the Fourier transforms
$\tilde{g}_i(\mathbf{k})$ as in Appendix~A. Then (cf.\ Eq.~\eqref{A1}),
\begin{equation*}\label{B1}
g_j(\mathbf{x},t)
=\frac{\exp(i\mathbf{px})}{(2\pi)^{3/2}}\int d^3k\tilde{S}(\mathbf{k})
     \exp\left[-i\left(\mathbf{kx}+t\sqrt{|\mathbf{p}-\mathbf{k}|^2+m_j^2}\right)\right]
\end{equation*}
and we can rewrite Eq.~\eqref{I_12} as
\begin{gather}
I_{12}(t) =
\int d^3k\left|\tilde{S}(\mathbf{k})\right|^2 \nonumber \\
\times\exp\left\{it\left[\sqrt{\left(\mathbf{p}-\mathbf{k}\right)^2-m_1^2}-
\sqrt{\left(\mathbf{p}-\mathbf{k}\right)^2-m_2^2}\right]\right\}.
\label{B2}
\end{gather}
The square roots in Eq.~\eqref{B2} prevents exact calculation of the integral.
However it can be estimated for the ultrarelativistic case and  taking into account
that the function $\tilde{S}(\mathbf{k})$ is only appreciable for
$|\mathbf{k}|\ll|\mathbf{p}|$. These simplifications allow us to use the approximation
\begin{equation}\label{B3}
\sqrt{\left(\mathbf{p}-\mathbf{k}\right)^2-m_1^2}-
\sqrt{\left(\mathbf{p}-\mathbf{k}\right)^2-m_2^2}
\approx \frac{m_1^2-m_2^2}{2|\mathbf{p}|}
\left[1+\frac{\left(\mathbf{kp}\right)}{|\mathbf{p}|^2}\right].
\end{equation}
Substituting Eq.~\eqref{B3} into Eq.~\eqref{B2} we arrive at
\begin{equation*}\label{B4}
I_{12}=
\exp\left[-\frac{it(m_1^2-m_2^2)}{2|\mathbf{p}|}\right]
\int d^3k\left|\tilde{S}(\mathbf{k})\right|^2
\exp\left[-\frac{it(m_1^2-m_2^2)\left(\mathbf{kp}\right)}{2|\mathbf{p}|^3}\right].
\end{equation*}
In particular, for the Gaussian distribution given by Eq.~\eqref{A4} and
considering that 
\begin{equation*}\label{B5}
I_{12}(0) = \int d^3x \left|g(\mathbf{x})\right|^2=1,
\end{equation*}
we obtain
\begin{align*}\label{B6}
I_{12}(t)
& = \exp\left[-\frac{it(m_1^2-m_2^2)}{2|\mathbf{p}|}
    -\left(\frac{|\mathbf{p}|t(m_1^2-m_2^2)}{4\sigma}\right)^2\right] \nonumber \\
& = \exp\left[-\frac{2i\pi t}{T_{\text{osc}}}
    -\left(\frac{t}{T_{\text{osc}}}\right)^2
     \left(\frac{\pi}{|\mathbf{p}|\sigma}\right)^2\right],
\end{align*}
where
\[
T_{\text{osc}}=\frac{4\pi|\mathbf{p}|}{m_1^2-m_2^2}.
\]


\end{document}